\documentclass[aps,preprint,superscriptaddress,groupedaddress,nofootinbib]{revtex4}  
\usepackage[pdftex]{graphicx,color}
\usepackage{amsmath, amssymb}
\usepackage{color}
\usepackage{siunitx}
\usepackage{bm}
\usepackage{here}
\usepackage{url}
\usepackage{comment}
\usepackage{bigints}
\usepackage{braket}
\usepackage{mhchem}
\usepackage{lscape}
\usepackage{wrapfig}
\usepackage{cases}
\usepackage{ulem}
\usepackage{cancel}
\usepackage{cases}

{\list{}{%
\rightmargin=1zw%
\leftmargin=1zw%
}\item\relax}{\endlist}

\newcommand{\Space} {\:\:\:\:\:\:\:\:\:\:\:}
\newcommand{\sspace} {\:\:\:\:\:}
\newcommand{\Beq} {\begin{equation} }
\newcommand{\Eeq} {\end{equation} }
\newcommand{\beq} {\begin{equation*} }
\newcommand{\eeq} {\end{equation*} }
\newcommand{\Beqa} {\begin{eqnarray} }
\newcommand{\Eeqa} {\end{eqnarray} }
\newcommand{\beqa} {\begin{eqnarray*} }
\newcommand{\eeqa} {\end{eqnarray*} }

\newcommand{\meter} {\si{\meter}}

\newcommand{\second} {\si{\second} }

\newcommand{\Arccos}{ \mathrm{Arccos} }

\begin{document}

\preprint{KOBE-COSMO-18-10, MAD-TH-18-10}

\title{Searching for Bispectrum of Stochastic Gravitational Waves\\
 with Pulsar Timing Arrays}

\author{Makoto Tsuneto}
\email[]{makoto.tsuneto@stonybrook.edu}
\affiliation{SUNY Stony Brook Department of Physics and Astronomy, Stony Brook 11794-3800, USA}

\author{Asuka Ito}
\email[]{asuka-ito@stu.kobe-u.ac.jp}

\affiliation{Department of Physics, Kobe University, Kobe 657-8501, Japan}

\author{Toshifumi Noumi}
\email[]{tnoumi@phys.sci.kobe-u.ac.jp}

\affiliation{Department of Physics, Kobe University, Kobe 657-8501, Japan}

\affiliation{Department of Physics, University of Wisconsin-Madison, Madison, WI 53706, USA}

\author{Jiro Soda}
\email[]{jiro@phys.sci.kobe-u.ac.jp}

\affiliation{Department of Physics, Kobe University, Kobe 657-8501, Japan}

\date{\today}

\begin{abstract}
We study how to probe bispectra of stochastic gravitational waves
with pulsar timing arrays.
The bispectrum is a key to probe the origin of stochastic gravitational waves.
In particular, the shape of the bispectrum carries valuable information of inflation models.
We show that an appropriate filter function for three point correlations 
enables us to extract a specific configuration of momentum triangles in bispectra.
We also calculate the overlap reduction functions and discuss strategy for detecting
the bispectrum with multiple pulsars.
\end{abstract}

\maketitle

\tableofcontents

\section{Introduction}
Stochastic gravitational waves (GWs), the GW analog for
the cosmic microwave background, are going through us from all directions.
They contain primordial GWs~\cite{Grishchuk:1974ny},\cite{Starobinsky:1979ty} produced during 
inflation~\cite{Starobinsky:1980te}-\cite{Linde:1981mu} 
in addition to GWs of cosmological/astrophysical origin~\cite{Regimbau:2011rp}.
As the name suggests, stochastic GWs are characterized by statistics such as 
the power spectrum, bispectrum and  higher order correlation functions.

The detection of the power spectrum of stochastic GWs is a clue to probe the early universe.
Moreover, the bispectrum of stochastic GWs, which represents the non-gaussianity,
is a powerful tool to discriminate astrophysical and primordial origin since 
the former has a gaussian distribution as long as event rates are 
high enough to create continuous GWs~\cite{Regimbau:2011rp},\cite{Coward:2006df}.%
\footnote{
Alternatively, if the event rate is too low to produce 
continuous GWs, the distribution is not gaussian. 
Detectability of such feature is explored in~\cite{Drasco:2002yd}-\cite{Thrane:2013kb}.
}
Therefore, the bispectrum of stochastic GWs enables us to probe the early universe.
Indeed, the bispectrum of primordial GWs contains 
the detail of inflation models like nonlinear interactions of the graviton.
The shape of bispectra, which depends on
inflation models~\cite{Baumann:2009ds}-\cite{Bartolo:2004if}, 
allows us to discriminate inflation models.%
\footnote{
Recently, it was shown that 
the detection of power spectrum of primordial GWs is not enough 
to exclude bouncing universe models~\cite{Ito:2016fqp}.
Therefore, the bispectrum is also important to distinguish inflation and bouncing universe models.%
}
Furthermore, the imprint of new particles 
with the mass comparable to the Hubble scale during inflation can potentially appear
in the squeezed limit of momentum 
triangles~\cite{Dimastrogiovanni:2018gkl}-\cite{Lee:2016vti}.
Therefore, the bispectrum is a powerful probe of the early universe and 
beyond the standard model.

Now, GW detectors are in operation to probe stochastic GWs, although 
no signal of stochastic GWs has been detected yet.
The sensitive frequency band of interferometers like LIGO~\cite{TheLIGOScientific:2016dpb} and Virgo~\cite{Abadie:2011fx} is around $10^{2}$ Hz,
while pulsar timing arrays such as EPTA~\cite{Lentati:2015qwp} and 
 NANOGrav~\cite{Arzoumanian:2018saf} are searching for stochastic GWs with 
a frequency range $10^{-9}$-$10^{-7}$ Hz.
The constraints on the energy density of stochastic GWs are $\Omega_{GW} < 1.2 \times 10^{-9}$ (EPTA),  
$\Omega_{GW} < 3.4 \times 10^{-10}$ (NANOGrav), and $\Omega_{GW} < 1.1 \times 10^{-11}$ (PPTA), respectively.
In future, the space interferometers,
 LISA~\cite{Bartolo:2016ami} and  DECIGO~\cite{Sato:2017dkf}, 
will be launched in a few decades. 
The pulsar timing array project SKA~\cite{Janssen:2014dka} will start 
in 2020 and significantly improve the current sensitivity.
Its possible upper limit is 
$\Omega_{GW} < 1.0 \times 10^{-13}$~\cite{Zhao:2013bba}.
Therefore, it is worth exploring a new theoretical research area 
for forthcoming observations.

In this paper, we investigate a method for detecting the bispectrum
of stochastic GWs with pulsar timing arrays.
We utilize a filter function not only to maximize the signal to noise ratio (SNR),
but also to extract a specific configuration of momentum triangles in the bispectrum.
We generalize the overlap reduction function (ORF) in the two point 
correlation function~\cite{Allen:1997ad}-\cite{Romano:2016dpx} to three point correlation functions.
The result is useful for increasing the sensitivity 
with correlation of multiple pulsars~\cite{Lentati:2015qwp},\cite{Arzoumanian:2018saf}. 
Moreover, optimal pulsar configurations for each polarization mode of bispectra will be found.

The paper is organized as follows.
In section II we explain how stochastic GWs affect the residual of pulsar timing.
In section III we review the detection method for the power spectrum of 
stochastic GWs.
There, we obtain the Hellings-Downs curve~\cite{Hellings:1983fr}.
In section IV we extend the discussion to the case of the bispectrum.
A certain filter function is chosen to detect the bispectral shape 
of stochastic GWs.
In section V we explain a method for calculating ORFs which depends on 
the pulsar configuration and the momentum triangle that we would like to probe. 
In section VI we analyze ORFs for (+++) mode. 
In section VII we examine ORFs for (+$\times \times$) mode. 
In section VIII  we discuss the ORF for circularly polarized modes.
The final section is devoted to the conclusion.
\section{GW signal in pulsar timing arrays}
In the Minkowski spacetime, GWs as tensor perturbations of the metric can be 
expanded with plane waves:
\Beq
h_{ij}(t, \vec{x}) = \sum_A \, \int_{-\infty}^{\infty} {\mathrm d} f \, \int {\mathrm d} \hat{\Omega} \, 
e^{  2\pi {\mathrm i}   \left( ft - |f| \hat{\Omega}\cdot \vec{x} \right) } \, 
\tilde{h}_A (f, \hat{\Omega}) e_{ij}^A (\hat{\Omega}) \,,
\label{eq: expansion}
\Eeq
where 
$\hat{\Omega}$ is the direction of propagation of GWs.
Polarization tensors, which satisfy 
$e_{ij}^{A}(\hat{\Omega}) e_{ij}^{A'}(\hat{\Omega}) = 2 \delta^{AA'}$, can be defined by
\Beq
e_{ij}^+(\hat{\Omega}) = \hat{m}_i \hat{m}_j - \hat{n}_i \hat{n}_j \ , \quad		
e_{ij}^{\times}(\hat{\Omega}) = \hat{m}_i \hat{n}_j + \hat{n}_i \hat{m}_j \ .
\label{eq: eplus}
\Eeq
Here, $\hat{m}$ and $\hat{n}$ are unit vectors
perpendicular to $\hat{\Omega}$ and one another. 
It should be noted that since the way to choose the directions of $\hat{m}$ and $\hat{n}$ is arbitrary and thus 
the linear polarization bases (\ref{eq: eplus}) depend on coordinates, circular polarization bases,
\begin{equation}
  e^{{\rm R}}_{ij}(\hat{\Omega}) = \frac{e_{ij}^{+}(\hat{\Omega}) + i e_{ij}^{\times}(\hat{\Omega})}
                                  {\sqrt{2}} \  , \quad
  e^{{\rm L}}_{ij}(\hat{\Omega}) = \frac{e_{ij}^{+}(\hat{\Omega}) - i e_{ij}^{\times}(\hat{\Omega})}
                                  {\sqrt{2}} \  ,
\end{equation}
are physically essential. However, from now on, we employ $+$ and $\times$ polarizations for convenience in calculation until Sec.\,\ref{circu}. 

GW detectors have their specific response to GWs. 
For instance, pulsars can be utilized as a detector.
A pulsar is a neutron star which emits periodic electromagnetic fields
very accurately.
If gravitational waves $h_A (f, \hat{\Omega})$ exist continuously
between the Earth and a pulsar (the direction $\hat{p}$), 
we observe the redshift of an emitted pulse as \cite{Anholm:2008wy}
\Beq
\tilde{Z}(f, \hat{\Omega}) 
          = \left( e^{ - 2\pi {\mathrm i} L \left( f + |f| \hat{\Omega}\cdot \hat{p}  
                         \right) }  - 1 \right) 
            \sum_A \tilde{h}_A (f, \hat{\Omega}) F^A( \hat{\Omega},\hat{p} ) \ ,
           \label{z}
\Eeq
where 
\Beq
F^A(\hat{\Omega},\hat{p}) \equiv e_{ij}^A \, \cfrac{1}{2} \, \cfrac{ \hat{p}^i \hat{p}^j }{ 1 + \hat{\Omega}\cdot \hat{p} }  
\label{eq: pattern}
\Eeq
is the pattern function.
It represents a geometrical factor, namely, dependence of the sensitivity
on the configuration of the detector and GWs.
Furthermore, we integrate Eq.\,(\ref{z})
\Beq
\tilde{z}(f) = \int {\mathrm d} \hat{\Omega} \, \tilde{Z} (f, \hat{\Omega} ) \  ,
\label{zz}
\Eeq
because stochastic GWs propagate toward all directions.
The quantity that is actually measured is the residual defined by
\Beqa
R(t) &\equiv&  \int_0^t {\mathrm d}t' \int_{-\infty}^{\infty} {\mathrm d}f \, 
e^{2\pi {\mathrm i} f t' }	\, \tilde{z}(f)	
\\
&=& \int_0^t {\mathrm d}t' \, \int_{-\infty}^{\infty} {\mathrm d}f \, \int {\mathrm d} \hat{\Omega} \, 
e^{2\pi {\mathrm i} f t' }	\, 	
\left( e^{ - 2\pi {\mathrm i} L \left( f + |f| \hat{\Omega}\cdot \hat{p} \right) }  - 1 \right) \, 
\sum_A \tilde{h}_A (f, \hat{\Omega}) F^A( \hat{\Omega},\hat{p} )		\ .
\label{resi}
\Eeqa
Since, for pulsar timing measurement, the minimum frequency is about $0.1 \,\mathrm{yr^{-1}}$ and the shortest distance between the Earth and a pulsar
is $\sim 100 \,\mathrm{ly}$, we have $fL \gtrsim 10$. 
In this range, the exponential term in the parenthesis of Eq.\,(\ref{resi}) can be approximated to zero 
because it oscillates rapidly. 
Hence, Eq.\,(\ref{zz}) can be approximated as 
\Beq
\tilde{z}(f) \simeq
    - \sum_A \int {\mathrm d} \hat{\Omega}
       \tilde{h}_A (f, \hat{\Omega}) F^A( \hat{\Omega},\hat{p} ) \ .
       \label{zzz}
\Eeq
We find that the correlation of residuals is directly related 
with that of stochastic GWs.
Therefore, observing appropriate correlation function of the signals, 
we can probe the statistic of stochastic GWs.
\section{Probing power spectrum with pulsar timing arrays} \label{sectwo}
In practice, output data of detectors $s_{i}$ ({\it i} is the index of detectors)
contain noises $n_i$ peculiar to each detector in addition to GW signals. 
We can remove the noises 
by correlating multiple detectors.
In this section, we review the detection method of power spectra of stochastic GWs with 
two detectors \cite{Maggiore:1999vm}.
In pulsar timing measurement, the signal is 
\Beq
s_i (t) = z_i (t) + n_i (t)		\,,
\Eeq
where we have assumed linearity of the signal.
Consider $n_i$ are zero-mean random noises and irrelevant to each other. 
Also, we assume there is no correlation between GW signals and noises.
We then have
\Beqa
\braket{n_i(t)} &=& 0			\,,	
\label{eq: ni}
\\
\braket{n_i(t) n_j(t)} &=& 0		\,,	\sspace	({\rm for} \ i \neq j)		
\label{eq: nn}
\\
\braket{n_i(t) z_j(t)} &=& 0			\,.
\label{eq: nz}
\Eeqa
Now, we correlate two output signals of two pulsars (the directions $\hat{p}_{1}$ and $\hat{p}_{2}$) as
\Beq
S_{12} \equiv \iint_{-T/2}^{T/2} {\mathrm d} t\,  {\mathrm d}t'\, s_1(t)s_2(t') Q(t, t')			\,,
\label{eq: S12}
\Eeq
where $T$ is the observation time and
$Q(t, t')$ is a filter function which specifies the way to correlate $s_1(t)$ and $s_2(t)$. 
We take a filter function as $Q(t-t')$ to extract contributions from the stochastic GWs
which satisfy the momentum conservation.
In the frequency domain, Eq.\,(\ref{eq: S12}) is expanded as
\Beq
S_{12} = \iint_{-T/2}^{T/2} \mathrm{d}t  \mathrm{d}t' \, 
           \iiint_{-\infty}^{\infty} \mathrm{d}f \mathrm{d}f' \mathrm{d}f''
             \tilde{s}_1(f) \, \tilde{s}_2(f') \tilde{Q}(f'')
              e^{2\pi \mathrm{i} ft} e^{2\pi \mathrm{i} f't'}
              e^{2\pi \mathrm{i} f''(t-t')}  \ .
\label{eq: FourierS12}
\Eeq
Thus, if T is large enough to suffice $ fT, f'T \gg 1$, the ranges of integrals are approximated as $T \rightarrow \infty$. Then, Eq.\,(\ref{eq: FourierS12}) is approximated to
\Beq
S_{12} = \iint_{-\infty}^{\infty} {\mathrm d}f {\mathrm d}f'\, \delta (f + f') \tilde{s}_1(f) \tilde{s}_2(f') \tilde{Q}(f')   \ .
\label{20}
\Eeq
We note that the appropriate functional form of $\tilde{Q}(f)$ is determined 
in such a way that the SNR is maximized
(see Appendix.\ref{apn}).
From Eqs.\,(\ref{eq: nn}),(\ref{eq: nz}) and (\ref{20}) the ensemble average
of the correlation $S_{12}$ is
\Beq
\braket{ S_{12} } = \iint_{-\infty}^{\infty} {\mathrm d} f {\mathrm d}f' 
\delta( f + f' ) \braket{ \tilde{z}_1(f) \tilde{z}_2(f') } \tilde{ Q }(f') \ .  \label{mumu}
\Eeq

Here let us define the spectral density of GWs as
\Beq
\braket{ \tilde{h}_A(f, \hat{\Omega}) \tilde{h}_{A'} (f', \hat{\Omega '}) } 
  = \frac{1}{4\pi} \, \delta ( \hat{\Omega} + \hat{\Omega'} ) \delta( f + f' )
     \delta_{A A'}  \frac{1}{2} \, S_h (|f|)	\,,
\label{eq: unpolarized}
\Eeq
where we have assumed that 
there is no polarization of the GW.
The delta functions come from the fact that 
the Minkowski spacetime is homogeneous and the stochastic GW is stationary.
From Eqs.\,(\ref{zzz}),(\ref{mumu}) and (\ref{eq: unpolarized}), we obtain
\Beq
\braket{ S_{12} } = 2 T \, \int_{0}^{\infty} {\mathrm d}f \, \frac{1}{2} S_h(f) \tilde{Q}(f) \Gamma  \ ,  \label{munosiki}
\Eeq
where we used $  \delta(0) = \int_{-T/2}^{T/2} {\mathrm d}t  = T$ and
\Beq
\Gamma = \sum_A \, \int {\mathrm d} \hat{\Omega} \, \frac{1}{4\pi} \, 
                    F^A (\hat{\Omega},\hat{p}_{1}) F^A (-\hat{\Omega},\hat{p}_{2})			\,.
\label{eq: Gamma(f)}
\Eeq
It includes the pattern function $F^A (\hat{\Omega},\hat{p_{i}})$ defined by Eq.\,(\ref{eq: pattern}) and
represents the loss of sensitivity due to relative directions of pulsars and called the overlap reduction function (ORF).
One can perform the angular integral of (\ref{eq: Gamma(f)}) and the result is 
\Beq
\Gamma = \frac{1}{3} +  \frac{1 + \cos \xi }{ 2 } 
         \left[ \ln \left( \frac{ 1 + \cos \xi }{ 2 } \right) - \frac{1}{6} \right]	\,,	
         \label{eq: Hellings-Downs}
\Eeq
where $\xi$ is the angle between $\hat{p}_{1}$ and $\hat{p}_{2}$.
From Eqs.\,(\ref{munosiki}) and (\ref{eq: Hellings-Downs}), 
the geometrical factor of $\braket{ S_{12} }$ depends only on $\xi$.
This is known as the Hellings-Downs curve~\cite{Hellings:1983fr}.%
\footnote{
Eq.\,(\ref{eq: Hellings-Downs}) differs from that of \cite{Hellings:1983fr}
because the definition of the two point correlation (\ref{eq: S12}) is different.
They correspond to each other by the exchange of $\xi \leftrightarrow (\pi - \xi)$.
}
It is characterized in that the ORF has quadrupolar signature and 
takes the maximum value when the two pulsars are in the opposite direction.
We find that the knowledge of the ORF is essential to extract $S_{h}(f)$ from the 
observable $\braket{ S_{12} }$.
Moreover, the ORF (\ref{eq: Hellings-Downs}) is useful for removing noises and increasing the sensitivity 
of two point correlations with multiple pulsars~\cite{Lentati:2015qwp},\cite{Arzoumanian:2018saf}.
Therefore, the ORF is a powerful tool in pulsar timing measurement.
This is true even in the case of three point correlations as we will see in the next section.
\section{Probing bispectrum with pulsar timing arrays} \label{secthree}
Let us extend previous discussion to three point correlations of GW signals
to probe the bispectrum of stochastic GWs.

First, we define the bispectrum as 
\begin{align}
\nonumber
&\braket{ \tilde{h}_A (f_1, \hat{\Omega}_1) \tilde{h}_{A'}(f_2, \hat{\Omega}_2) 
          \tilde{h}_{A''}(f_3, \hat{\Omega}_3) } 
\\
       &\quad= B_{AA'A''} (|f_1| , |f_2| , |f_3| ) \delta(f_{1}+f_{2}+f_{3})
\delta^{(3)}( |f_1| \hat{\Omega}_1 + |f_2| \hat{\Omega}_2 + |f_3| \hat{\Omega}_3) \  .
\label{eq: unpolarized3}
\end{align}
The first delta function denotes that correlations of stochastic GWs are time-independent. 
The second delta function shows that
the momenta form a closed triangle due to homogeneity of the Minkowski spacetime.
The bispectral shape varies depending on the inflation model, so that 
its measurement is a key to probe the early universe.

As in the previous section, 
correlation of three detectors can be defined with a filter function as follows:
\Beq
S_{123} = \iiint_{-T/2}^{T/2} \, \mathrm{d}t_1 \mathrm{d}t_2 \mathrm{d}t_3 \,\, s_1(t_1) s_2(t_2) s_3(t_3) Q(t_1, t_2, t_3)	\,.
\Eeq
We introduce a filter function in a form $Q(t_1, t_2, t_3) = Q(at_1 + bt_2 + ct_3)$, 
where $a,b$ and $c$ are positive constants. 
It will turn out that the filter function enables us to extract 
a specific configuration of the momentum triangle determined by these constants. 
Moving on to Fourier space, we have
\Beqa
S_{123} = \iiint_{-T/2}^{T/2} \mathrm{d}t_1 \mathrm{d}t_2 \mathrm{d}t_3 \, 
\iiiint_{-\infty}^{\infty} \mathrm{d}f_1 \mathrm{d}f_2 \mathrm{d}f_3 \mathrm{d}f \,\, \tilde{s}_1(f_1) \tilde{s}_2(f_2) \tilde{s}_3(f_3) \tilde{Q}(f)			
\nonumber
\\
\times e^{2\pi \mathrm{i} f_1 t_1} \, e^{2\pi \mathrm{i} f_2 t_2} \, e^{2\pi \mathrm{i} f_3 t_3} \, 
e^{-2\pi \mathrm{i} f (a t_1 + b t_2 + c t_3) } \ ,
\Eeqa
where we have assumed $\tilde{Q}(-f) = \tilde{Q}(f)$.
As in the case of the power spectrum,
taking $T \rightarrow \infty$, one can carry out the integration:
\Beqa
S_{123} &=& \iiiint_{-\infty}^{\infty} \mathrm{d}f_1  \mathrm{d}f_2  \mathrm{d}f_3  \mathrm{d}f \,
\tilde{s}_1(f_1) \tilde{s}_2(f_2) \tilde{s}_3(f_3) \tilde{Q}(f) \delta (f_1 - af)  \delta (f_2 - bf)  \delta (f_3 - cf)	
\nonumber
\\
&=&  \int_{-\infty}^{\infty} \mathrm{d}f \, \tilde{s}_1(af) \tilde{s}_2(bf) \tilde{s}_3(cf) \tilde{Q}(f) 		\,.
\label{eq: S123}
\Eeqa
Using Eqs.\,(\ref{eq: nn}) and (\ref{eq: nz}) in Eq.\,(\ref{eq: S123}), 
the ensemble average of $S_{123}$ becomes
\Beq
\braket{ S_{123} } =  \, \int_{-\infty}^{\infty} \mathrm{d}f \, \braket{ \tilde{z}_1(af) \tilde{z}_2(bf) \tilde{z}_3(cf) } \tilde{Q}(f)   \ .
\label{eq: mu3}
\Eeq
It should be noted that Eq.\,(\ref{eq: mu3}) is valid even for 
a single pulsar case as long as the noise is gaussian, namely, 
$\braket{n_i(t) n_i(t) n_{i}(t)} = 0$.
An explicit functional form of $\tilde{Q}(f)$ maximizing the SNR is also discussed in 
Appendix.\ref{apn}.
From Eqs.\,(\ref{zzz}), (\ref{eq: unpolarized3}) and (\ref{eq: mu3}),
one can deduce
\begin{eqnarray}
\braket{ S_{123} } = 2 T \sum_{A, A', A''} \int_{0}^{\infty} &\mathrm{d}f& 
                     \frac{1}{f^{3}} B_{AA'A''}(af, bf, cf) \tilde{Q}(f) 
                     \frac{\sin\big(\pi ( a+b+c ) f T \big)}{\pi ( a+b+c ) fT}  \nonumber \\
                &\times&    \frac{(4 \pi)^2}{abc} 
     \Gamma^{A A' A''} (a,b,c; \hat{p}_1,  \hat{p}_2,  \hat{p}_3)  \ ,
     \label{S123}
\end{eqnarray}
where the ORF for the three point correlation is defined by\,%
\footnote{
Note that we have defined the ORF to be scale invariant with respect to $a$,$b$ and $c$.}
\Beqa
  \Gamma ^{ A A' A'' } (a,b,c; \hat{p}_1, \hat{p}_2, \hat{p}_3) 
   &=& - \frac{abc}{(4\pi)^2}
       \iiint \mathrm{d}\hat{\Omega}_1  \mathrm{d}\hat{\Omega}_2 
                 \mathrm{d}\hat{\Omega}_3  
                 \delta^{(3)}( a \hat{\Omega}_{1} + b \hat{\Omega}_{2} + c \hat{\Omega}_{3} )
       \nonumber \\
       && \times 
       F^{A}(\hat{\Omega}_{1},\hat{p}_{1}) 
       F^{A'}(\hat{\Omega}_{2},\hat{p}_{2}) F^{A''}(\hat{\Omega}_{3},\hat{p}_{3}) \ . 
       \label{eq: ORF3}
\Eeqa
In Eq.\,(\ref{S123}), 
we see that the frequency and the angular integrals are separated due to the filter function,
although there appears a suppression factor, 
$\int_{-T/2}^{T/2} e^{2\pi i(a+b+c)ft} {\mathrm d}t 
= \frac{\sin\big(\pi ( a+b+c ) f T \big)}{\pi ( a+b+c ) f}$, in the first line.%
\footnote{
Allowing $a,b$ and $c$ to be negative, one can remove the suppression factor when 
$a+b+c=0$, i.e., in the collinear limit.
It implies that pulsar timing arrays is more sensitive to collinear limits of the bispectrum.
However, we only focus on positive constants case to probe general momentum triangles
in this paper.
}
Eq.\,(\ref{eq: ORF3}) shows that a specific configuration of the momentum triangle determined 
by $a,b$ and $c$ is extracted.%
\footnote{In an equal-time three point correlation function, all momentum triangles are integrated.
Such case is well studied  in the context of LISA~\cite{Bartolo:2018qqn}-\cite{Bartolo:2018rku}.
}

Therefore, we can probe the shape of the bispectrum 
with changing those parameters.
For this purpose, we need to carry out the angular integration to evaluate the ORF,
whose analytic form is available only for several special cases (see Appendix~\ref{anaana}).
For more general cases, we compute the ORF numerically in general cases instead.
The detail of the procedure will be explained in the next section.
\section{Overlap reduction function (ORF)} \label{apn2}
\begin{figure}[t]
\centering
\includegraphics[width=9cm,clip]{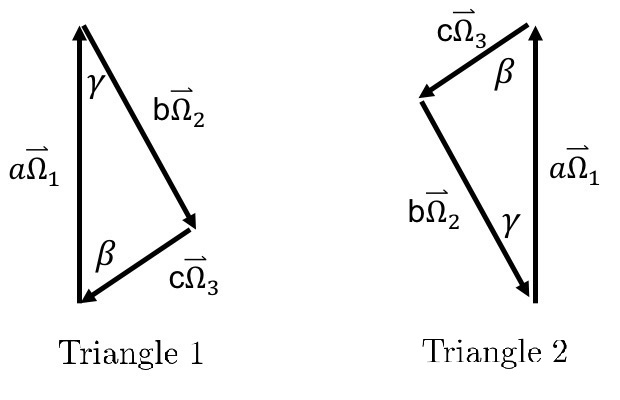}
\caption{Two types of triangles which satisfy the momentum conservation are depicted. 
The angles $\beta$ and $\gamma$ are related to $a,b,c$ as 
$\beta = \Arccos \left( \frac{ c^2 + a^2 - b^2 }{ 2 c a } \right) ({\rm for} \ c, a \neq 0)$ and
$\gamma = \Arccos \left( \frac{ a^2 + b^2 - c^2 }{ 2 a b } \right) ({\rm for} \ a, b \neq 0)$.
}
\label{fig: triangles}
\end{figure}

To treat the delta function in Eq.\,(\ref{eq: ORF3}) adequately, 
we use new integration variables which specify the rotation of momentum triangles.
There are two triangles that suffice 
$a \hat{\Omega}_1 + b \hat{\Omega}_2 + c \hat{\Omega}_3$ = 0 
as drawn in Fig.\,\ref{fig: triangles}, where $\beta$ and $ \gamma$ are angles which face with 
$b\hat{\Omega}_2$ and $c\hat{\Omega}_3$, respectively.
Since the shape of a triangle has already been determined by $(\beta, \gamma$), 
there remain three degrees of freedom regarding the rotation of the triangles.
They can be specified as follows:
\begin{enumerate}
\item the zenith angle of $\hat{\Omega}_1$ with respect to $z$-axis	: $\phi$
\item the azimuth angle of $\hat{\Omega}_1$ around the $z$-axis	: $\theta$
\item rotation of a triangle around $\hat{\Omega}_1$			: $\Phi$
\end{enumerate}
We will carry out the integral about these variables.

On the other hand,
the directions of pulsars from the Earth can be defined by
\Beq
\hat{p}_1 = \left( \begin{array}{ccc} 0\\ 0\\ 1 \end{array} \right) ,\sspace \hat{p}_2 = \left( \begin{array}{ccc}  \cos \xi_2 \sin \psi_2\\  \sin \xi_2 \sin \psi_2\\  \cos \psi_ 2 \end{array} \right) , \sspace  \hat{p}_3 = \left( \begin{array}{ccc}  \cos \xi_3 \sin \psi_3\\  \sin \xi_3 \sin \psi_3\\  \cos \psi_ 3 \end{array} \right) \  .
\label{ppp}
\Eeq
Note that we have fixed $\hat{p}_{1}$ to be the $z$-direction.
Although one can set $\xi_{3} = 0$ without loss of generality, we leave it to keep $\xi_{2}$ and $\xi_{3}$ symmetric for a while.

Below we express $F^{A}(\hat{\Omega}_i,\hat{p}_{i})$ defined by Eq.\,(\ref{eq: pattern}) in terms of
the constants $\gamma, \beta, \xi_{2}, \psi_{2}, \xi_{3}, \psi_{3}$ and the integration variables 
$\phi, \theta, \Phi$ to carry out the integral (\ref{eq: ORF3}).
\subsection{Computation of Pattern functions}
The direction of $\hat{\Omega}_{1}$ is
\Beq
 \hat{\Omega}_1 = R_z(\phi) \, R_y(\theta) \, \vec{e}_z =  \left( \begin{array}{ccc}  \cos \phi \sin \theta\\  \sin \phi \sin \theta\\  \cos \theta\end{array} \right)		\,,
 \label{eq: GW1}
\Eeq
where $R_z(\phi)$ and $R_y(\theta)$ represent
the rotation around $z$-axis by $\phi$ and the rotation around $y$-axis by $\theta$, respectively.

To compute $e_{ij}^A(\hat{\Omega}_1)$, it is necessary to determine 
$\hat{m}(\hat{\Omega}_1)$ and  $\hat{n}(\hat{\Omega}_1)$. 
Since any $\hat{m}(\hat{\Omega}_1)$ and  $\hat{n}(\hat{\Omega}_1)$ are allowed 
as long as $\hat{m}(\hat{\Omega}_1)$ and $\hat{n}(\hat{\Omega}_1)$ are perpendicular 
to $\hat{\Omega}_1$ and one another, 
we choose the following simple ones:
\Beq
\hat{m} (\hat{\Omega}_1) = R_z(\phi) \, R_y(\theta) \, \vec{e}_x = \left( \begin{array}{ccc}  \cos \phi \cos \theta \\ \sin \phi \cos \theta \\ -\sin \theta \end{array} \right)		,\sspace \hat{n} (\hat{\Omega}_1) =  R_z(\phi) \, R_y(\theta) \, \vec{e}_y = \left( \begin{array}{ccc}  -\sin \phi \\ \cos \phi \\ 0 \end{array} \right)	.
\label{eq: mn}
\Eeq
Then, from Eqs.\,(\ref{eq: eplus}) and (\ref{eq: pattern}), 
the pattern functions for $\hat{\Omega}_1$ follow as
\begin{eqnarray}
  \begin{cases}
F^{+} (\hat{\Omega}_1,\hat{p}_{1})  = \frac{1}{2} (1-\cos\theta) \ ,  & \\
F^{\times} (\hat{\Omega}_1,\hat{p}_{1}) = 0 \  .  &
  \end{cases} 
  \label{eq: F1}
\end{eqnarray}
Next we compute $F^{A'}(\hat{\Omega}_2,\hat{p}_{2})$ and 
$F^{A''}(\hat{\Omega}_3,\hat{p}_{3})$. 
Let us focus on the Triangle 1 for a while.
Then, the zenith angle of $\hat{\Omega}_2$ with respect to $\hat{\Omega}_1$ is $\pi - \gamma$ and the azimuth angle of $\hat{\Omega}_2$ around the $\hat{\Omega}_1$ is $\Phi$. Similarly, the zenith angle of $\hat{\Omega}_3$ with respect to $\hat{\Omega}_1$ is $\pi - \beta$ and the azimuth angle of $\hat{\Omega}_3$ around the $\hat{\Omega}_1$ is 
$\Phi + \pi$.
Therefore, we have
\begin{eqnarray}
\begin{cases}
\hat{\Omega}_2 = R_z(X_2) \, R_y(Y_2) \vec{e}_z = R_z(\phi) \, R_y(\theta) \, R_z(\Phi) \,  R_y(\pi - \gamma) \, R_y(-\theta) \, R_z(-\phi) \hat{ \Omega }_1 \ ,  & \\
\hat{\Omega}_3 = R_z(X_3) \, R_y(Y_3) \vec{e}_z = R_z(\phi) \, R_y(\theta) \, R_z(\Phi+\pi) \,  R_y(\pi - \beta) \, R_y(-\theta) \, R_z(-\phi) \hat{ \Omega }_1  \ ,
  &
\end{cases} 
\label{eq: GW3}
\end{eqnarray}
where we have defined the zenith angles $X_{2} , X_{3}$ and the azimuth angles 
$Y_{2}, Y_{3}$ with respect to z-axis for later use.
More explicitly, they obey the equations
\Beq
\left\{ \,\,
\begin{aligned}
\left( \begin{array}{ccc} \cos X_2 \sin Y_2 \\ \sin X_2 \sin Y_2 \\  \cos Y_2  \\ \end{array} \right) &=& 
\left( \begin{array}{ccc} \cos \Phi \cos \theta \sin \gamma \cos \phi - \cos \phi \sin \theta \cos \gamma - \sin \gamma \sin \Phi \sin \phi \\ \cos \Phi \cos \theta \sin \gamma \sin \phi - \sin \phi \sin \theta \cos \gamma + \sin \gamma \cos \phi \sin \Phi \\ -\sin \theta \cos \Phi \sin \gamma - \cos \theta \cos \gamma \\ \end{array} \right)		\,,	\\
\left( \begin{array}{ccc} \cos X_3 \sin Y_3 \\ \sin X_3 \sin Y_3 \\  \cos Y_3  \\ \end{array} \right) &=& 
\left( \begin{array}{ccc}  -\cos \Phi \cos \theta \cos \phi \sin \beta - \cos \phi \sin \theta \cos \beta + \sin \Phi \sin \phi \sin \beta \\ -\cos \Phi \cos \theta \sin \phi \sin \beta - \sin \phi \sin \theta \cos \beta - \cos \phi \sin \Phi \sin \beta\\ \sin \theta \cos \Phi \sin \beta - \cos \theta \cos \beta \\\end{array} \right)	\,.
\end{aligned}
\right.
\Eeq
We can solve the above equations to obtain 
\Beq
Y_2 = \mathrm{Arccos} 
\left( - \sin \theta \cos \Phi  \sin \gamma  - \cos \theta  \cos \gamma \right)  \ ,	
\label{eq: Y2}
\Eeq
\Beq
X_2 = 
\left\{
\begin{aligned}
0	\Space\Space\Space\Space (Y_2 = 0, \pi)		\\
\,\,	\\
\mathrm{Arccos}  \left[  \cfrac{ \cos \Phi \cos \theta \sin \gamma \cos \phi - \cos \phi \sin \theta \cos \gamma - \sin \gamma \sin \Phi \sin \phi }{ \sqrt{ 1 - \left( \sin \theta \cos \Phi \sin \gamma + \cos \gamma \cos \theta \right)^2 } } \right]
\\     		 ( \cos \Phi \cos \theta \sin \gamma \sin \phi - \sin \phi \sin \theta \cos \gamma + \sin \gamma \cos \phi \sin \Phi \geq 0 ) 		\\
\,\,	\\
2 \pi - \mathrm{Arccos}  \left[  \cfrac{ \cos \Phi \cos \theta \sin \gamma \cos \phi - \cos \phi \sin \theta \cos \gamma - \sin \gamma \sin \Phi \sin \phi }{ \sqrt{ 1 - \left( \sin \theta \cos \Phi \sin \gamma + \cos \gamma \cos \theta \right)^2 } } \right]
\\	({\rm otherwise})
\end{aligned}
\right.   
\label{eq: X2}
\Eeq
and
\Beq
Y_3 = \Arccos( \sin \theta \cos \Phi \sin \beta - \cos \theta \cos \beta )  \ ,
\Eeq
\Beq
X_3 =
\left\{
\begin{aligned}
0	\Space\Space\Space\Space (Y_3 = 0, \pi)		\\
\,\,	\\
\mathrm{Arccos} \left[ \cfrac{ - \cos \Phi \cos \theta \cos \phi \sin \beta - \cos \phi \sin \theta \cos \beta + \sin \Phi \sin \phi \sin \beta }{ \sqrt{ 1 - \left( \sin \theta \cos \Phi \sin \beta - \cos \theta \cos \beta \right)^2  } } \right]		\\
(-\cos \Phi \cos \theta \sin \phi \sin \beta - \sin \phi \sin \theta \cos \beta - \cos \phi \sin \Phi \sin \beta \geq 0 )			\\
\,\,	\\
2 \pi - \mathrm{Arccos} \left[ \cfrac{ - \cos \Phi \cos \theta \cos \phi \sin \beta - \cos \phi \sin \theta \cos \beta + \sin \Phi \sin \phi \sin \beta }{ \sqrt{ 1 - \left( \sin \theta \cos \Phi \sin \beta - \cos \theta \cos \beta \right)^2  } } \right]	
\\	({\rm otherwise})
\end{aligned}
\right.  \label{eq: X3}
\Eeq

On the other hand, unlike $\hat{\Omega}_i$, $\hat{m}(\hat{\Omega}_i)$ and  
$\hat{n}(\hat{\Omega}_i)$ depend on the process of rotation rather than the entire rotation. 
Therefore, $\hat{m}(\hat{\Omega}_{2}), \hat{n}(\hat{\Omega}_{2}), 
\hat{m}(\hat{\Omega}_{3})$ and $\hat{n}(\hat{\Omega}_{3})$
have to follow the same rotation process of $\hat{m}(\hat{\Omega}_{1})$ and 
$\hat{n}(\hat{\Omega}_{1})$.
Thus,
\Beq
\left\{ \,\, 
\begin{aligned}
\hat{m}(\hat{ \Omega }_2) &=& R_z (X_2) R_y(Y_2) \vec{e}_x 		\\
\hat{n}(\hat{ \Omega }_2) &=& R_z (X_2) R_y(Y_2) \vec{e}_y 
\end{aligned}
\right.
\ \ \  , \sspace \sspace
\left\{ \,\, 
\begin{aligned}
\hat{m}(\hat{ \Omega }_3) &=& R_z (X_3) R_y(Y_3) \vec{e}_x 		\\
\hat{n}(\hat{ \Omega }_3) &=& R_z (X_3) R_y(Y_3) \vec{e}_y 
\end{aligned}
\right.  \ \ . \label{mnmn}
\Eeq
Now, we can express $F^{A'}(\hat{\Omega}_2,\hat{p}_{2})$ and $F^{A''}(\hat{\Omega}_3,\hat{p}_{3})$ by
the constants $\gamma, \beta, \xi_{2}, \psi_{2}, \xi_{3}, \psi_{3}$ and the integration variables 
$\phi, \theta, \Phi$ through Eqs.\,(\ref{eq: eplus}), (\ref{eq: pattern}),
(\ref{ppp}), (\ref{eq: GW3}) and (\ref{eq: Y2})-(\ref{mnmn}).
\subsection{ORF for ($+++$) and ($+\times\times$) modes}
For example, the ORF for (+++) mode is evaluated as
\Beq
\Gamma^{+++} (\gamma, \beta; \hat{p}_1, \hat{p}_2, \hat{p}_3) 
   = - \cfrac{1}{ (4 \pi)^2} 
\, \int_0^{2 \pi} {\mathrm d} \phi \, \int_{-1}^{1} {\mathrm d} \cos \theta \, \int_0^{2 \pi} {\mathrm d} \Phi \, \Big[ \ f(\gamma, \hat{p}_2 ; \beta, \hat{p}_3 ) + 
         f( \beta, \hat{p}_3 ; \gamma, \hat{p}_2 ) \  \Big]  \ ,
\label{eq: Gamma+++}
\Eeq
where
\Beq
f( \gamma, \hat{p}_2; \beta, \hat{p}_3 )
= \frac{1}{8} \left( 1 - \cos \theta \right) \, \prod_{i=2, 3} \left[ \frac{ \left( \hat{p}_i \cdot \hat{m}(\hat{\Omega}_i) \right)^2 - \left(  \hat{p}_i \cdot \hat{n}(\hat{\Omega}_i) \right)^2 }{ 1 + \hat{p}_i \cdot \hat{\Omega}_i } \right] \,		\,.
\label{eq: Triangle1}
\Eeq
In Eq.\,(\ref{eq: Gamma+++}), the first and the second terms in the square bracket are from
the Triangle 1 and the Triangle 2 of Fig.\,\ref{fig: triangles}, respectively.
We can not calculate (\ref{eq: Gamma+++}) analytically in general.
However, in a few special cases, such as the co-aligned pulsars/a single pulsar
($\hat{p}_1 = \hat{p}_2 = \hat{p}_3$) and
the anti-parallel pulsars/two oppositely directed pulsars 
($\hat{p}_1 = - \hat{p}_2 = -\hat{p}_3$), 
the integral can be solved analytically.
The analytical expressions are given in Appendix.\ref{anaana}.

Taking a look at Eq.\,(\ref{eq: F1}), we find that the ORF for 
($\times A'A''$) are all zero. 
Then, there just remain ($++\times$) and ($+\times\times$) modes.
However, we found that the integrand for the ($++\times$) mode is odd function,
so that the ORF is zero. 
This can be interpreted as follows: In calculating the ORF for the three point correlation, we sum over all possible triangles.
As shown in Fig.\,\ref{fig: triangles}, this corresponds to adding parity-inverted triangles.
Since the response to cross polarization, namely, $F^{\times} (\hat{\Omega}_{i}, \hat{p}_{i})$ ($i=2,3$), has odd parity, the contributions from polarization combinations where cross appears only once should be zero.
Consequently, we have only ($+++$) and ($+\times\times$) modes which have nonzero values.
In the next section, we will show the several results of the ORF 
for ($+++$) and ($+\times\times$) modes.
\section{ORF for ($+++$) mode} \label{pppppp}
\begin{figure}[H]
\centering
\includegraphics[width=9cm,clip]{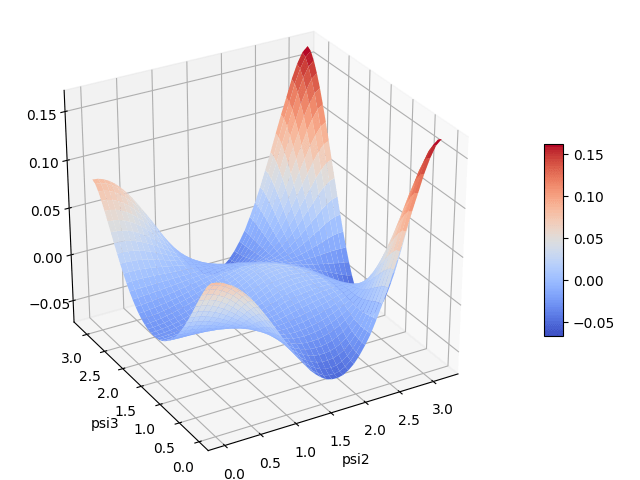}
\caption{$(-\Gamma^{+++})$ for the squeezed triangle against 
$\psi_2$ and $\psi_3$ ($ \xi_2 = 0 $) is depicted.}
\label{fig: SQ_1}
\end{figure}

\begin{figure}[H]
\centering
\includegraphics[width=9cm,clip]{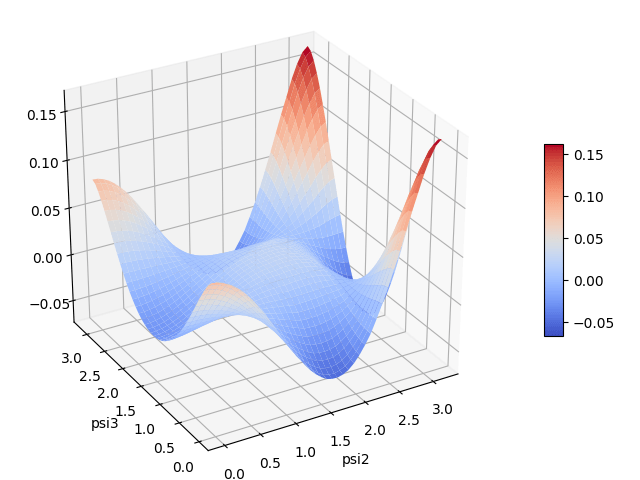}
\caption{$(-\Gamma^{+++})$ for the squeezed triangle against 
$\psi_2$ and $\psi_3$ ($ \xi_2 = \frac{\pi}{2} $) is depicted.}
\label{fig: SQ_2}
\end{figure}

\begin{figure}[H]
\centering
\includegraphics[width=9cm,clip]{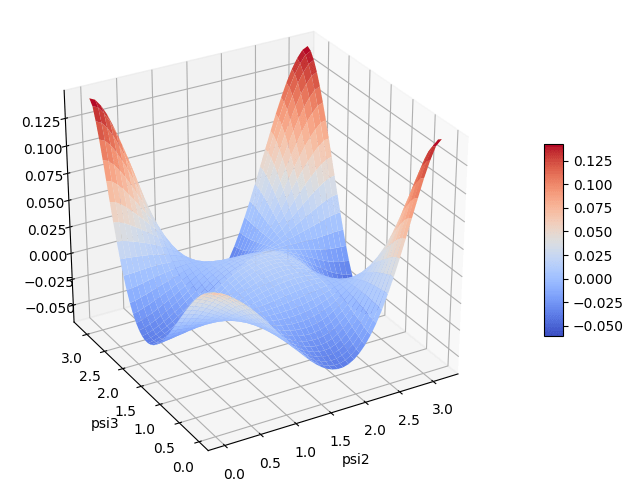}
\caption{$(-\Gamma^{+++})$ for the equilateral triangle against 
$\psi_2$ and $\psi_3$ ($ \xi_2 = 0 $) is depicted.}
\label{fig: EQ_1}
\end{figure}

\begin{figure}[H]
\centering
\includegraphics[width=9cm,clip]{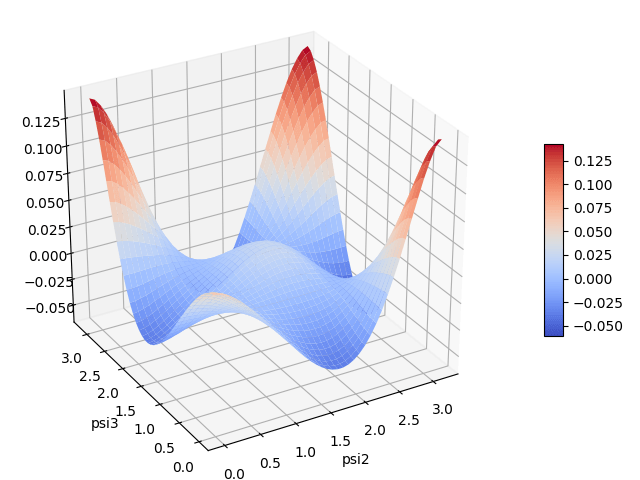}
\caption{$(-\Gamma^{+++})$ for the equilateral triangle against 
$\psi_2$ and $\psi_3$ ($ \xi_2 = \frac{\pi}{2} $) is depicted.}
\label{fig: EQ_2}
\end{figure}

In this section, we evaluate the ORF for $(A,A',A'')=(+,+,+)$ mode 
while changing parameters $(a,b,c,\hat{p}_1, \hat{p}_2, \hat{p}_3)$.
Let us show the dependence of the ORF on pulsar configurations in 
the squeezed and the equilateral momentum triangles.
We have calculated the ORF dependence on $\psi_2$ and $\psi_3$, for 
$\xi_2 = 0, \pi/20, 2\pi/20, 3\pi/20, \ldots, 39\pi/20$. 
The definition of the angles $\xi_{i}$ and $\psi_{i}$ are given in Eq.\,(\ref{ppp}) and
we have set $\xi_{3}=0$ without loss of generality.
Since it turns out that change in $\xi_2$ does not drastically affect 
the behavior of the ORF, the ORFs for $\xi_2 = 0, \frac{\pi}{2}$ 
are shown in Figs.\,\ref{fig: SQ_1}-\ref{fig: EQ_2} as illustrative examples.

Consider the squeezed momentum triangle: $\frac{b}{a}=1, \frac{c}{a} \rightarrow 0$.
In Figs.\,\ref{fig: SQ_1} and \ref{fig: SQ_2}, we plotted the ORF for the squeezed momentum triangle.
From Figs.\,\ref{fig: SQ_1} and \ref{fig: SQ_2}, we see extremums are on the four edges, 
$(\psi_2, \psi_3) = (0, 0), (0, \pi), (\pi, 0), (\pi, \pi)$, and the central point $(\psi_2, \psi_3) = (\pi/2, \pi/2)$.
In particular, the ORFs are maximized when 
$(\psi_2, \psi_3) = (\pi, 0), (\pi, \pi)$, namely, 
\begin{equation}
\hat{p}_1 = -\hat{p}_2 = \hat{p}_3	\ ,\sspace	\mbox{or}	\sspace	
-\hat{p}_1 = \hat{p}_2 = \hat{p}_3  \ .
\end{equation}
Notice that $\hat{p}_1 = \hat{p}_2 = -\hat{p}_3$ does not have the maximum value
since $\hat{p}_{1,2}$ and $\hat{p}_{3}$ are not symmetric due to the choice $a=b, c=0$.
We also see quadrupolar signature along with $\psi_{2}$ and $\psi_{3}$ directions.
It is analogous to the Hellings-Downs curve.

Next, we consider the equilateral momentum triangle: $\frac{b}{a}=\frac{c}{a}=1$.
Figs.\,\ref{fig: EQ_1} and \ref{fig: EQ_2} show the ORF for this case.
At first sight,   
Figs.\,\ref{fig: EQ_1} and \ref{fig: EQ_2} are similar to 
the squeezed triangle case, Figs.\,\ref{fig: SQ_1} and \ref{fig: SQ_2}.
In fact, the positions of the extremums are the same. 
However, Figs.\,\ref{fig: SQ_3} and \ref{fig: SQ_4} have 
maximum points at 
$(\psi_2, \psi_3) = (\pi, 0), (\pi, \pi), (\pi, 0)$, namely, 
\begin{equation}
\hat{p}_1 = -\hat{p}_2 = \hat{p}_3	\ , \sspace	\mbox{or}	\sspace	
-\hat{p}_1 = \hat{p}_2 = \hat{p}_3  \ , \sspace	\mbox{or}	\sspace
\hat{p}_1 = \hat{p}_2 = -\hat{p}_3  \ . \label{opop}
\end{equation}
At these maximum points, there is a permutation symmetry 
of three pulsars. 
It is because o$a=b=c$.
Quadrupolar signature along with $\psi_{2}$ and $\psi_{3}$ directions appears in 
Figs.\,\ref{fig: EQ_1} and \ref{fig: EQ_2} as well as Figs.\,\ref{fig: SQ_1} and \ref{fig: SQ_2}.
We confirmed that such signature is general for any momentum triangle.
Furthermore, ORFs take maximum values at the anti-parallel pulsar configuration for 
any momentum triangle.
Therefore, the anti-parallel pulsar configuration is optimal for the detection of ($+++$) mode
of the bispectrum.
\section{ORF for ($+\times\times$) mode} \label{pcc}
\begin{figure}[H]
\centering
\includegraphics[width=9cm,clip]{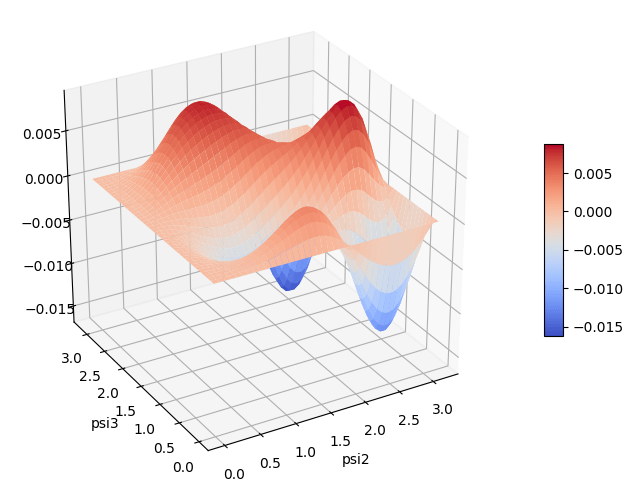}
\caption{$(-\Gamma^{+\times\times})$ for the squeezed triangle against 
$\psi_2$ and $\psi_3$ ($ \xi_2 = 0 $) is depicted.
There is a maximum of the absolute value at $\psi_{2}=2.6, \psi_{3}=2.6$.}
\label{fig: SQ_3}
\end{figure}

\begin{figure}[H]
\centering
\includegraphics[width=9cm,clip]{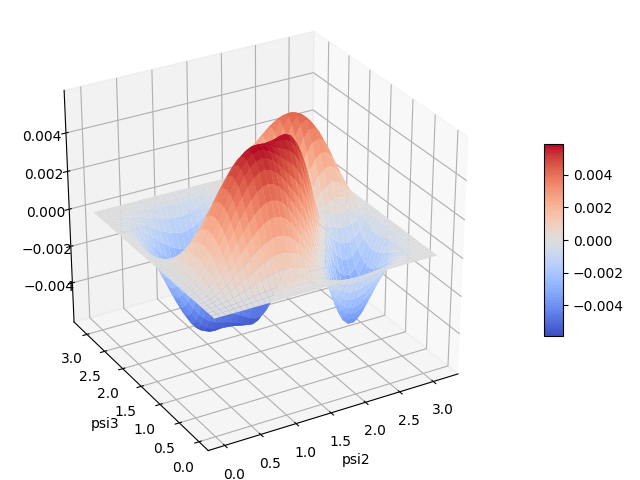}
\caption{$(-\Gamma^{+\times\times})$ for the squeezed triangle against 
$\psi_2$ and $\psi_3$ ($ \xi_2 = \frac{\pi}{2} $) is depicted.
There are maximums of the absolute value at $\psi_{2}=1.1, \psi_{3}=0.94$ and $\psi_{2}=1.1, \psi_{3}=2.2$.}
\label{fig: SQ_4}
\end{figure}

\begin{figure}[H]
\centering
\includegraphics[width=9cm,clip]{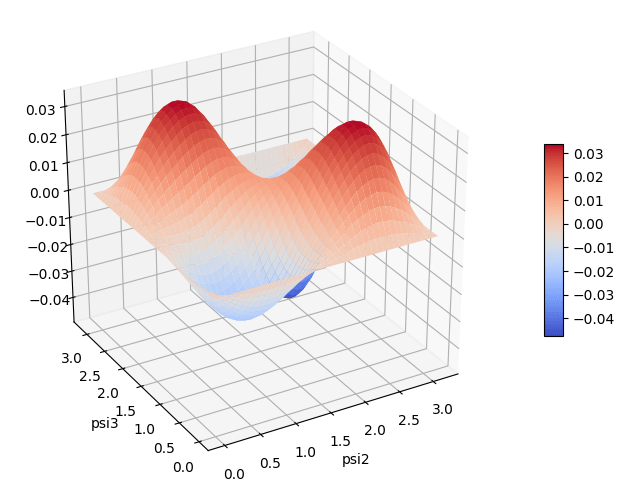}
\caption{$(-\Gamma^{+\times\times})$ for the equilateral triangle against 
$\psi_2$ and $\psi_3$ ($ \xi_2 = 0 $) is depicted.
There is a maximum of the absolute value at $\psi_{2}=2.4, \psi_{3}=2.4$.}
\label{fig: EQ_3}
\end{figure}

\begin{figure}[H]
\centering
\includegraphics[width=9cm,clip]{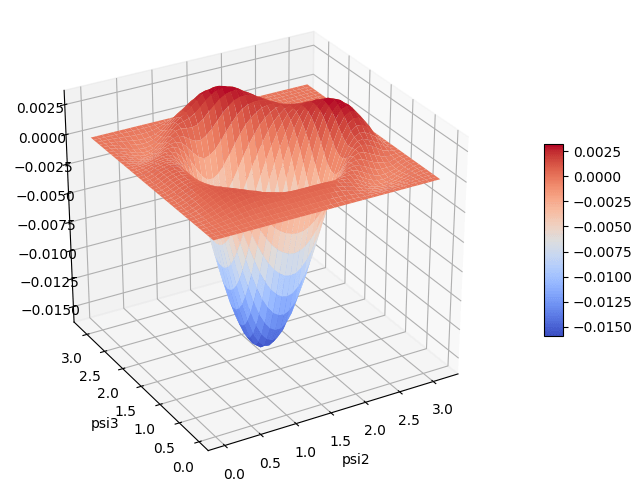}
\caption{$(-\Gamma^{+\times\times})$ for the equilateral triangle against 
$\psi_2$ and $\psi_3$ ($ \xi_2 = \frac{\pi}{2} $) is depicted.
There is a maximum of the absolute value at $\psi_{2}=1.5, \psi_{3}=1.5$.}
\label{fig: EQ_4}
\end{figure}

As discussed in section \ref{apn2}, the ORFs $\Gamma^{AA'A''}$, are zero other than 
$\Gamma^{+++}$ and $\Gamma^{+\times\times}$.
Thus, in this section, we study $\Gamma^{+\times\times}$.
Figs.\,\ref{fig: SQ_3} and \ref{fig: SQ_4} show the dependence of the ORF
on the pulsar configurations for the squeezed triangle case.
Figs.\,\ref{fig: EQ_3} and \ref{fig: EQ_4} are for the equilateral triangle case.
From  Figs.\,\ref{fig: SQ_3}-\ref{fig: EQ_4},
one can see that $\Gamma^{+\times\times}$ does not have sensitivity 
to the co-aligned and the anti-parallel pulsar configurations. 
Therefore, the co-aligned and the anti-parallel pulsar configurations 
are suitable to probe $(+++)$ mode of the bispectrum since 
then only $\Gamma^{+++}$ is non-zero.

On the other hand, we have to choose pulsar configurations other than
the co-aligned and the anti-parallel pulsar configurations 
to probe $(+\times\times)$ mode of the bispectrum.
Then one can always find an optimal pulsar configuration corresponding to each momentum triangle.
For example, the optimal pulsar configuration for the case of Fig.\,\ref{fig: SQ_3} is 
$\psi_{2}=2.6 , \psi_{3}=2.6$.
Potentially, we can separate ($+++$) and ($+\times\times$) modes 
in (\ref{S123}) with multiple pulsars because 
$\Gamma^{+++}$ and $\Gamma^{+\times\times}$ are complementary, i.e., 
positions of peaks are different as depicted in Figs.\,\ref{fig: SQ_1}-\ref{fig: EQ_4}. 
\section{ORF for circular polarization} \label{circu}
So far, we have used the linear polarization tensors 
$e_{ij}^{+}$ and $e_{ij}^{\times}$ as bases of GWs.
In this subsection, we investigate the ORF for circular polarization bases defined by
\begin{equation}
  e^{{\rm R}}_{ij}(\hat{\Omega}) = \frac{e_{ij}^{+}(\hat{\Omega}) + i e_{ij}^{\times}(\hat{\Omega})}
                                  {\sqrt{2}} \  , \quad
  e^{{\rm L}}_{ij}(\hat{\Omega}) = \frac{e_{ij}^{+}(\hat{\Omega}) - i e_{ij}^{\times}(\hat{\Omega})}
                                  {\sqrt{2}} \  .  \label{cirbase}
\end{equation}
They then satisfy
\begin{equation}
  \left( e^{\sigma}_{ij}(\hat{\Omega}) \right)^{*} e^{\sigma'}_{ij}(\hat{\Omega}) 
    = 2 \delta^{\sigma\sigma'} \ ,  \quad
  \left( e^{{\rm R}}_{ij}(\hat{\Omega}) \right)^{*} 
    = e^{{\rm R}}_{ij}(-\hat{\Omega}) 
    = e^{{\rm L}}_{ij}(\hat{\Omega}) \ ,
\end{equation}
where $\sigma,\sigma' = {\rm R},{\rm L}$.
The circular polarization bases are useful to discuss the signature of parity 
violation~\cite{Soda:2011am}-\cite{Bartolo:2017szm}.
From Eqs.\,(\ref{eq: pattern}), (\ref{eq: ORF3}) and (\ref{cirbase}), it is easy to construct the ORF 
for the circular polarization bases as 
\begin{eqnarray}
\begin{cases}
\    \Gamma^{{\rm RRR}} = \Gamma^{{\rm RLL}} = \Gamma^{{\rm LLL}} = \Gamma^{{\rm LRR}}
               =      \frac{1}{2\sqrt{2}} 
                 \big[ \Gamma^{+++} - \Gamma^{+\times\times}  \big] \ ,   &  \\
\    \Gamma^{{\rm RRL}} = \Gamma^{{\rm RLR}} = \Gamma^{{\rm LLR}} = \Gamma^{{\rm LRL}}
               =      \frac{1}{2\sqrt{2}} 
                 \big[ \Gamma^{+++} + \Gamma^{+\times\times}  \big] \ .  &
\label{licir}
\end{cases} 
\end{eqnarray}
They can be evaluated immediately by using the result in previous subsections.
Especially, for the case of the co-aligned (or a single pulsar)
and the anti-parallel (or two oppositely directed pulsars) configurations, 
all the ORF for circular polarization bases are given by Eqs.\,(\ref{exact1}) and (\ref{exact2})
because then $\Gamma^{+\times\times}$ is zero. 

We find that the bispectrum can not be used to detect circular polarization since
all the ORFs (\ref{licir}) are parity even and thus 
all parity violating bispectrum terms such as 
$\left( B_{{\rm RRR}}-B_{{\rm LLL}} \right) \left( \Gamma^{{\rm RRR}} - \Gamma^{{\rm LLL}} \right)$
are zero.
The situation is same for the case of the power spectrum~\cite{Kato:2015bye}.
Physically, 
the three point correlation function is on a plane rather than three-dimensional
because we have ignored the pulsar term in Eq.\,(\ref{z}), and thus
the three point function can not distinguish left and right polarizations.
Therefore, one guess that the four point correlation function could detect parity violation with pulsar timing arrays.
\section{Conclusion}
We explored the possibility of the detection of the bispectrum of stochastic gravitational waves with pulsar timing arrays.
We showed that an appropriate filter function in three point correlations enables 
us to extract a specific configuration of momentum triangles in the bispectrum of stochastic GWs.
Therefore, one can probe the bispectral shape, which carries important information of 
the early universe, by adjusting the filter function.

Furthermore, we considered the three point correlation in pulsar timing measurement and
investigated the dependence of the sensitivity, i.e. ORF, on
the pulsar configurations and the momentum triangle which we want to probe.
Several examples of the result are shown in Figs.\,\ref{fig: SQ_1}-\ref{fig: EQ_4}.
In ($+++$) mode of the ORF, it was found that the anti-parallel (or two oppositely directed pulsars)
configuration universally maximizes the sensitivity of the three point correlation.
This is analogous to the ORF for the two point correlation (\ref{eq: Hellings-Downs}) where
the configuration for two oppositely directed pulsars has a maximum value.
Moreover, as is shown in Figs.\,\ref{fig: SQ_1}-\ref{fig: EQ_2}, ($+++$) mode has quadrupolar signature
like the Hellings-Downs curve (\ref{eq: Hellings-Downs}).
Remarkably, we obtained analytical expressions of the ORF for special cases,
(\ref{exact1}) and (\ref{exact2}). 
In ($+\times\times$) mode, the ORF does not have the sensitivity in 
the co-aligned and the anti-parallel pulsar configurations as is depicted in 
Figs.\,\ref{fig: SQ_3}-\ref{fig: EQ_4}.
Therefore, the co-aligned and the anti-parallel pulsar configurations 
are suitable to probe $(+++)$ mode of the bispectrum.
Conversely, we have to choose pulsar configurations other than
the co-aligned and the anti-parallel pulsar configurations 
to probe $(+\times\times)$ mode of the bispectrum.
Then one can always find an optimal pulsar configuration corresponding to each momentum configuration.
Potentially, we can separate ($+++$) and ($+\times\times$) modes 
in (\ref{S123}) with multiple pulsars because $\Gamma^{+++}$ and $\Gamma^{+\times\times}$
are complementary, i.e., 
positions of peaks are different as depicted in Figs.\,\ref{fig: SQ_1}-\ref{fig: EQ_4}. 
The ORF for circular polarization bases are also discussed in section \ref{circu} 
and described in Eq.\,(\ref{licir}).
We note that our result, i.e. ORF, is useful for removing noises and increasing the sensitivity with correlation of 
multiple pulsars~\cite{Lentati:2015qwp},\cite{Arzoumanian:2018saf}.

We roughly estimated the detectable value of the non linear estimator $f_{NL}$ by setting the SNR (\ref{ssnnrr})
to be unity,
on the assumption that the observation time $T \sim 100$ years and 
target frequencies $f \sim 3\times 10^{-9}$ Hz, with 
$100$ pulsars which detectable timing residuals are $\sim 10$ ns.
We then got $f_{NL} \sim 10^{12}$.\,%
\footnote{
Inhomogeneity of the universe might supress the bispectrum of GWs during propagation~\cite{A.Lewis}.
Such effect is not accounted for in this estimate.
}
Then, one may wonder if the non-gaussianity is too small to detect at all~\cite{Maldacena:2002vr}.
However, there is another GW source rather than vacuum fluctuations such as 
particle production during inflation~\cite{Senatore:2011sp} and 
nonlinear density perturbations after inflation~\cite{Ananda:2006af}.
They can produce non-gaussian GWs.
For example, vector field production during inflation can induce abundant primordial GWs with large 
non-gaussianity~\cite{Cook:2013xea}.
Non-Abelian gauge fields can also generate very large tensor non-gaussianity 
because they have a tensor component~\cite{Agrawal:2017awz}.
Therefore, the detection of the bispecta is promising.
\begin{acknowledgments}
We would like to thank A.\,Lewis for helpful comments.
A.\,I. would like to thank N.\,Bartolo, S.\,Matarrese, E.\,Dimastrogiovanni, M.\,Fasiello and S.\,R\"as\"anen
for useful comments and discussions.
A.\,I. was supported by Grant-in-Aid for JSPS Research Fellow and JSPS KAKENHI Grant No.JP17J00216.
T.\,N. was supported in part by JSPS KAKENHI Grant Numbers JP17H02894 and JP18K13539, and
MEXT KAKENHI Grant Number JP18H04352. 
J.\,S. was in part supported by JSPS KAKENHI
Grant Numbers JP17H02894, JP17K18778, JP15H05895, JP17H06359, JP18H04589.
J.\,S and T.\,N are also supported by JSPS Bilateral Joint Research
Projects (JSPS-NRF collaboration) “String Axion Cosmology.”
\end{acknowledgments}
\appendix
\section{Optimum filter function} \label{apn}

In section \ref{sectwo} and \ref{secthree}, 
we derived the expression of the two point correlation (\ref{mumu}) and 
the three point correlation (\ref{S123}), respectively.
There we left the filter functions $\tilde{Q}(f)$ as 
an arbitrary function.
In this appendix we investigate which form of the filter function $\tilde{Q}(f)$ maximizes the SNR.

To begin with, the variance of $S_{12}$ is defined as
\Beq
\sigma^2 \equiv \braket{S_{12}^2} - \braket{ S_{12} }^2 		\,.
\Eeq
Using Eqs.\,(\ref{eq: nn}) and (\ref{eq: nz}), we obtain
\Beqa
\sigma^2 &=& \int_{-\infty}^{\infty} \mathrm{d}f \mathrm{d}f' \, \tilde{Q}(f) \tilde{Q}(f') 
\times \left[ \braket{ \tilde{s}_1 (f) \tilde{s}_2 (f) \tilde{s}_1(f') \tilde{s}_2(f') }-\braket{ \tilde{s}_1(f) \tilde{s}_2 (f) } \braket{ \tilde{s}_2(f') \tilde{s}_1 (f') } \right]
\nonumber \\
&\simeq& \int_{-\infty}^{\infty} {\mathrm d}f {\mathrm d}f' \, 
\tilde{Q}(f) \tilde{Q}(f')\, 
\braket{ \tilde{n}_1 (f) \tilde{n}_1 (f') } \braket{ \tilde{n}_2(f) \tilde{n}_2(f') }		\,.
\label{eq: variance}
\Eeqa
where we have assumed $z_i \ll n_i$.
Defining the spectral noise density 
\Beq
\braket{ \tilde{n}_i(f) \tilde{n}_j(f') } = \cfrac{1}{2}\, \delta(f + f') \, \delta_{ij} \, S_n^{(i)}(|f|)  \ ,
\label{eq: noise}
\Eeq
we can rewrite Eq.\,(\ref{eq: variance}) as
\Beq
\sigma^2 = \frac{T}{2} \int_{0}^{\infty} \mathrm{d}f \,  \tilde{ Q }(f)^2 \, S_n^{(1)}(f) S_n^{(2)}(f) 		\,.
\label{eq: sigma_2}
\Eeq
The SNR can be defined as $\frac{\braket{S_{12}}}{\sqrt{\sigma^{2}}}$ and then 
we have
\Beq
{\rm SNR} = \cfrac{\sqrt{2T} \, \int_{0}^{\infty} 
                  {\mathrm d}f \,  S_h(f) \tilde{Q}(f)\Gamma	}
{ \left( \int_{0}^{\infty} \mathrm{d}f \, \tilde{ Q }(f)^2 \, S_n^{(1)}(f) S_n^{(2)}(f) \right)^{1/2} } \ .
\label{snr}
\Eeq
From Eq.\,(\ref{snr}), the optimum $\tilde{Q}(f)$ which maximizes the SNR is 
\Beq
\tilde{Q}(f) \propto \cfrac{ S_h(f) }{ S_n^{(1)}(f) S_n^{(2)}(f) } \ .
\Eeq
It is determined by $S_h(f)$, $S_n^{(1)}(f)$ and $S_n^{(2)}(f)$.

Next, we extend above discussion to the case of the three point correlation.
One can obtain the variation of $S_{123}$ in the same manner as the two point 
correlation case:
\Beq
\sigma^2 = \frac{T^{2}}{4abc} \, 
\int_0^{\infty} \mathrm{d}f \tilde{Q}(f)^2 S_n^{(1)}(af) S_n^{(2)}(bf) S_n^{(3)}(cf)  \ .
\Eeq
Then, the SNR defined by $\frac{\braket{S_{123}}}{\sqrt{\sigma^{2}}}$ is
\Beq
{\rm SNR} = 
\frac{
  4 \sum_{A, A', A''}
  \int_0^{\infty} \mathrm{d}f \frac{1}{f^{3}} \, B_{AA'A''}(af, bf, cf) \tilde{Q}(f) 
   \frac{\sin\big(\pi ( a+b+c ) f T \big)}{\pi ( a+b+c ) fT} (4 \pi)^2
  \Gamma^{A A' A''} (a,b,c; \hat{p}_1,  \hat{p}_2,  \hat{p}_3) }
{ \left( abc \int_0^{\infty} \mathrm{d}f  \tilde{Q}(f)^2 S_n^{(1)}(af) S_n^{(2)}(bf) S_n^{(3)}(cf) \right)^{1/2}
 }   \ .  \label{ssnnrr}
\Eeq
We get the optimum filter function by requiring it to maximize the SNR as
\Beq
\tilde{Q}(f) \propto
   \cfrac{\sum_{A, A', A''} \frac{1}{f^{3}} B_{AA'A''}(af, bf, cf ) 
           \frac{\sin\big(\pi ( a+b+c ) f T \big)}{\pi ( a+b+c ) fT}
           \Gamma^{A A' A''} (a,b,c; \hat{p}_1,  \hat{p}_2,  \hat{p}_3)}
         { S_n^{(1)}(af)  S_n^{(2)}(bf)  S_n^{(3)}(cf) }  \ .
\Eeq
It is determined by the bispectrum, the ORF and the noises.
\section{Analytical expression of ORF for special cases} \label{anaana}
For the cases of the co-aligned pulsars/a single pulsar
($\hat{p}_1 = \hat{p}_2 = \hat{p}_3$) and
the anti-parallel pulsars/two oppositely directed pulsars 
($\hat{p}_1 = - \hat{p}_2 = -\hat{p}_3$), 
we can solve the integral (\ref{eq: Gamma+++}) analytically.
The results are
\begin{numcases}
  {}
  \Gamma^{+++} (\beta, \gamma) = -\frac{ 1 }{24} 
             \big( 3 + \cos (\beta - \gamma) - \cos\beta - \cos\gamma \big) \quad 
             ({\rm for \  co\mathchar`-aligned})  \ , \label{exact1} & \\
  \Gamma^{+++} (\beta, \gamma) = -\frac{ 1 }{24} 
             \big( 3 + \cos (\beta - \gamma) + \cos\beta + \cos\gamma \big) \quad 
             ({\rm for \  anti\mathchar`-parallel}) \ . \label{exact2}  &
\end{numcases}

In Eq.\,(\ref{exact1}),
there is a peak at $\beta=\gamma=\frac{\pi}{2}$, namely, the squeezed momentum triangle.
On the other hand, Eq.\,(\ref{exact2})
has a peak at $\beta=\gamma=0$, namely, the folded momentum triangle.
From Eqs.\,(\ref{exact1}) and (\ref{exact2}), we see that 
the ORF for the anti-parallel pulsar configuration is always larger than 
that for the co-aligned pulsar configuration.
We numerically confirmed that the anti-parallel pulsar configuration 
maximizes the ORF for any momentum triangle.
Therefore, the anti-parallel pulsar configuration is optimal for the detection of ($+++$) mode 
of the bispectrum.

We mention that numerical evaluations of the ORFs corresponded with the analytical 
solutions (\ref{exact1}) and (\ref{exact2}).
It justifies the numerical calculation of ORFs.

\end{document}